# Formation of quantum spin Hall state on Si surface and energy gap scaling with strength of spin orbit coupling

*Miao Zhou[1], Wenmei Ming[1], Zheng Liu[1], Zhengfei Wang[1], Yugui Yao[2], and Feng Liu[1,3*]*

[1] Department of Materials Science and Engineering, University of Utah, UT 84112

[2] School of Physics, Beijing Institute of Technology, Beijing, China 100081

[3] Collaborative Innovation Center of Quantum Matter, Beijing, China 100084

*fliu@eng.utah.edu*

For potential applications in spintronics and quantum computing, it is desirable to place a quantum spin Hall insulator [i.e., a 2D topological insulator (TI)] on a substrate while maintaining a large energy gap. Here, we demonstrate a unique approach to create the large-gap 2D TI state on a semiconductor surface, based on first-principles calculations and effective Hamiltonian analysis. We show that when heavy elements with strong spin orbit coupling (SOC) such as Bi and Pb atoms are deposited on a patterned H-Si(111) surface into a hexagonal lattice, they exhibit a 2D TI state with a large energy gap of $\geq 0.5$ eV. The TI state arises from an intriguing substrate orbital filtering effect that selects a suitable orbital composition around the Fermi level, so that the system can be matched onto a four-band effective model Hamiltonian. Furthermore, it is found that within this model, the SOC gap does not increase monotonically with the increasing strength of SOC. These interesting results may shed new light in future design and fabrication of large-gap topological quantum states.

Recently there has been a surge in the investigation of topological insulators (TIs) [1–3]. TIs are characterized by topologically protected metallic surface or edge states with helical spin polarization residing inside an insulating bulk gap. These states have negligible elastic scattering and Anderson localization [4, 5], which may provide ideal dissipationless spin current for future electronic devices with low power consumption. To realize their potential applications, it is desirable for the TIs to have an energy gap as large as possible [6], i.e., for room temperature applications. As for 2D TIs [i.e., quantum spin Hall (QSH) insulators], they also need to be grown or placed onto a substrate [7-11] or formed as an interface [12, 13], while maintaining a large gap. One desired approach is to directly fabricate large-gap TIs on semiconductor surfaces, which may avoid problems like transfer or interfacing a 2D layer over a foreign substrate [11]. So far, however, this goal remains allusive.

The HgTe quantum well, as the first theoretically predicted [12] and experimentally confirmed [12] QSH insulator, has a small gap of 40 meV with topological edge states only detectable at very low temperature (<10 K) [13]. Recent studies pertaining to Bi/Sb(111) films [14–17], Sn films [18], metal-decorated graphene [19–21], silicene/germanene [22] and 2D organometallic frameworks [23–26] have largely enriched the family of 2D TIs, and some of them have a large gap [14, 18, 21]. However, a critical drawback with most previous theoretical studies of 2D TIs is their reliance on the electronic and topological properties of freestanding films, whose existence can be in doubt. Even a freestanding film does exist, its properties are expected to be influenced by the underlying substrate in real applications [8-11].

Here, we demonstrate a unique approach of creating QSH state on a conventional semiconductor surface via depositing heavy elements with strong spin orbit coupling (SOC) onto a patterned Si(111) surface into a hexagonal lattice, which exhibit TI state with a large energy gap of ~ 0.5 eV. Here, the substrate plays a 'positive' role acting as an orbital filter to critically select the orbital composition around the Fermi level to realize nontrivial large-gap topological state [27]. Specifically, the surface system can be matched onto an effective four-band model Hamiltonian which captures the underlying physics. We depict a unified picture of energy gap as a function of SOC to achieve large-gap QSH state. Importantly, we found that it is not necessarily true to have a large gap with a heavier atom of larger SOC, a noteworthy point for future design of TIs.

We have performed density functional theory (DFT) based first-principles calculations of band structure and band topology of 2D hexagonal lattices of various metal atoms, including Bi, Pb, Sb, Sn, Ga, In and Tl, grown on a patterned H-saturated Si(111) surface. The detailed methodologies are presented in the Supplementary Information [28]. We will first discuss in detail the results of Bi and Pb, as representative examples, and leave the results of other metals for later discussion. Atomically flat H-Si(111) surface has been prepared for decades and is a widely-used substrate for epitaxial growth of ordered overlayers [29, 30]. The surface dangling bonds are passivated by H to avoid surface reconstruction. In order to form a hexagonal metal overlayer lattice, we propose a two-step fabrication process, as shown in Fig. 1. First, to create a desirable surface template pattern for metal growth, H atoms are selectively removed in hexagonal symmetry using scanning tunneling microscopy, as discussed in Refs. [31, 32]; Second, heavy metal atoms with large SOC can be deposited to grow or self-assemble into the exposed Si sites, as already demonstrated for other systems [33-35].

We found a very strong binding between the deposited metal atoms and the exposed Si atoms in the H-Si(111) surface, as evidenced by the calculated adsorption length [$d$, see Fig. 1(c)] of 2.68 Å and 2.75 Å for Bi and Pb, respectively. The high structural stability is also indicated by a large adsorption energy ($E_{ad}$), defined as $E_{ad} = E_{M @H-Si(111)} - (2E_M + E_{H-Si(111)}) + E_{H2}$, where $E_{M @H-Si(111)}$, $E_M$, $E_{H-Si(111)}$ and $E_{H2}$ denote the energy of Bi/Pb@H-Si(111), single metal atom, pristine H-Si(111) surface and $H_2$ gas molecule, respectively. The adsorption energy is found to be 3.88 eV and 3.92 eV for Bi and Pb, respectively, which are much larger than the binding energies of bulk Bi (2.18 eV) and Pb (2.03 eV) in the crystalline solid form, indicating high thermodynamic stability of the surface systems.

To examine the band topology of such surface structures, we first purposely exclude SOC from calculation. The resulting band structures for Bi and Pb@H-Si(111) are shown in Figs. 2 (a-b), respectively. For Bi@H-Si(111), there are two Dirac bands residing inside the bulk gap of Si with a Dirac point at $K$ point, which locates nearby the Fermi level [Fig. 2(a)]. Analysis of band composition further showed that the two Dirac bands consist mainly of $p_x$ orbitals of Bi atoms. Another dispersive band, consisting of $p_y$ orbitals of Bi, sits below the bulk conduction band edge of Si and touches the upper Dirac band of $p_x$ orbitals at $\Gamma$ point. The band structure of Pb@H-Si(111) is similar to the case of Bi, represented by two Dirac bands inside the Si gap; but the Dirac point sits ~0.8 eV above Fermi level and the upper Dirac band largely overlaps with the conduction band of Si [Fig. 2(b)]. Such different behaviors originate from the different valance electron configuration of Pb and Bi, e.g., [Xe].$4f^{14}.5d^{10}.6s^2.6p^2$ for Pb and [Xe].$4f^{14}.5d^{10}.6s^2.6p^3$ for Bi. With two electrons less in Pb@H-Si(111) per unit cell, the lower Dirac band becomes almost unoccupied compared to that of Bi@H-Si(111).

Next, we include SOC in calculation, and the resulting band structures of Bi and Pb@H-Si(111) are shown in Figs. 2(c) and (d), respectively. One sees that for Bi@H-Si(111), two Dirac bands are split apart; one large energy gap of ~ 0.7 eV opens at $K$ point, and another gap of ~ 0.5 eV opens at $\Gamma$ point, which is the global gap. The $p_y$ and the upper $p_x$ bands are also separated by SOC, with an indirect gap ~ 0.45 eV. We note that the spin degeneracy of these bands is lifted with most noticeable splitting at *K* point, which is due to the Rashba effect [36] induced by the broken spatial inversion symmetry. Although Rashba effect has shown to be detrimental to QSH phase [19], in our systems such extrinsic spin splitting is relatively small compared to the intrinsic SOC induced band gap [see Fig. 2(c)], suggesting the QSH state is robust against Rashba effect in our surface systems. It should also be noted that the SOC strength of Si is orders of magnitude smaller that Bi, making the SOC of Bi a decisive factor in opening the energy gap. Similarly, the SOC opens a gap at *K* point for Pb@H-Si(111), with the upper branch of Dirac bands moving completely into the Si conduction band, as shown in Fig. 2(d). Nevertheless, the energy splitting between the two Dirac bands is found to be around 0.65 eV at *K* point, and an effective SOC gap of ~0.54 eV could be counted by the energy difference between the Si conduction band minimum and the top of the Pb $p_x$ band. However, to truly make the Pb@H-Si(111) a 2D TI, n-type doping is needed to shift the Fermi level into the SOC gap, such as by Si substrate doping or electric gating.

The SOC-induced gap opening at the Dirac point in Bi@H-Si(111) and Pb@H-Si(111) indicates possible existence of 2D TI state. To check this, we calculated the topological edge states of Bi@H-Si(111) by the Wannier90 package [37]. Using DFT bands as input, we construct the maximally localized Wannier functions and fit a tight-binding Hamiltonian with these functions. Figure 3(a) shows the DFT and fitted band structures, which are in very good agreement. Then, the edge Green's function of a semi-infinite Bi@H-Si(111) is constructed and the local density of state (DOS) of Bi zigzag edge is calculated, as shown in Fig. 3(b). Clearly, one sees gapless edge states that connect the upper and lower band edge of the bulk gap, forming a 1D Dirac cone at the center of Brillouin zone (*Γ* point). This indicates that the Bi@H-Si(111) is a 2D TI with a large gap of ~ 0.5 eV.

To further confirm the above topological edge-state results, we also calculated $Z_2$ topology number. As the spatial inversion symmetry is broken in these systems, we used the method developed by Xiao et al. [22, 38]. In this method, $Z_2$ is calculated by considering the Berry gauge potential and Berry curvature associated with the Bloch wave functions, which does not require any specific point-group symmetry. Indeed, we found that $Z_2 =1$ for Bi@H-Si(111) (Fig. S1 in Supplementary Information), confirming the existence of QSH state in this surface. Assuming a shift of Fermi level above the lower branch of Dirac band, we also found $Z_2 =1$ for Pb@H-Si(111) (Fig. S2).

The physical origin of QSH state in Bi@H-Si(111) and Pb@H-Si(111) can be understood by a substrate orbital filtering effect as discussed recently in a related system of Bi on halogenated Si surfaces [27]. Figure 3(c) shows the partial density of states (DOS) of Bi@H-Si(111) around the Fermi level. It is seen that the $p_z$ orbital of Bi hybridizes strongly with the dangling bond of the exposed surface Si atom overlapping in the same energy range, which effectively removes the $p_z$ bands away from the Fermi level, leaving only the $p_x$ and $p_y$ orbitals. We also analyzed the charge density redistribution [see upper panel of Fig. 3(d)], which clearly shows that charge redistribution induced by Si surface mainly happens to the $p_z$ orbital of Bi, in a similar way to the saturation of Bi $p_z$ orbital by using hydrogen [lower panel of Fig. 3(d)]. It has been shown that the free-standing planar hexagonal lattice of Bi is a topologically trivial insulator with $Z_2 = 0$ (see Fig. S3). When it is placed onto the H-Si(111) surface or adsorbed with H, it becomes topologically nontrivial (Figs. S1 and S4). This originates from the intriguing orbital filtering effect imposed by the substrate or H saturation, which selectively remove the $p_z$ orbitals from the Bi lattice to realize the large-gap QSH phase.

Specifically, we can describe the Bi@H-Si(111) using a simplified ($p_x$, $p_y$) four-band model Hamiltonian in a hexagonal lattice as [24,39],

$$\hat{H}_{eff} = \begin{pmatrix} \varepsilon_0 & 0 & S_{xx} & S_{xy} \\ 0 & \varepsilon_0 & S_{xy} & S_{yy} \\ S_{xx}^* & S_{xy}^* & \varepsilon_0 & 0 \\ S_{xy}^* & S_{yy}^* & 0 & \varepsilon_0 \end{pmatrix} + \sigma_z \lambda_{so} \begin{pmatrix} 0 & -i & 0 & 0 \\ i & 0 & 0 & 0 \\ 0 & 0 & 0 & -i \\ 0 & 0 & i & 0 \end{pmatrix}, \quad (1)$$

in which $S_{xx} = V_{pp\sigma} + (\frac{1}{4}V_{pp\sigma} + \frac{3}{4}V_{pp\pi}) * (e^{ika1} + e^{ik(a1+a2)})$, $S_{xy} = \frac{3}{4}(V_{pp\sigma} + V_{pp\pi}) * (e^{ika1} + e^{ik(a1+a2)})$,

$S_{yy} = V_{pp\pi} + (\frac{3}{4}V_{pp\sigma} + \frac{1}{4}V_{pp\pi}) * (e^{ika1} + e^{ik(a1+a2)})$, $a_1, a_2$ is the lattice vector, $V_{pp\sigma}$ ($V_{pp\pi}$) is the Slater-Koster parameter [40], and $\sigma_Z = \pm 1$ is the spin eigenvalue.

Diagonalization of Eq. (1) in reciprocal space gives the band structures shown in Fig. 4, which shows typical four bands as a function of SOC strength. One sees that without SOC, this Hamiltonian produces two flat bands and two Dirac bands with a Dirac point formed at $K$ point and two quadratic points at $\Gamma$ point [Fig. 4(a)]. Inclusion of a small SOC ($\lambda = 0.2t$) opens one energy gap ($\Delta E_1$) at $K$ point and two energy gaps ($\Delta E_2$) at $\Gamma$ point [Fig. 4(b)], with both gaps topologically nontrivial [24]. With the increasing SOC strength, both $\Delta E_1$ and $\Delta E_2$ increase [Fig. 4(c)], which eventually leads to the formation of a different energy gap ($\Delta E_3$) between the upper and lower Dirac bands at $\Gamma$ point when $\Delta E_3$ becomes smaller than both $\Delta E_1$ and $\Delta E_2$ [Fig. 4(d)]. As such, for sufficiently large SOC, $\Delta E_3$ replaces $\Delta E_1$ to be the global gap, and correspondingly the global gap shifts from $K$ to $\Gamma$ point. Further increase of SOC will tend to decrease $\Delta E_3$, indicating that for sufficiently large SOC the band gap decreases with increasing SOC.

Such interesting phenomenon has also been confirmed by the DFT results. By comparing Bi@H-Si(111) and Pb@H-Si(111), we see that given the correct Fermi energy, the global gap is located at $\Gamma$ point for Bi@H-Si(111) [Fig. 2(c)], but at $K$ point for Pb@H-Si(111) [Fig. 2(d)]. This is because the SOC strength in $p$ orbital of Pb (~0.91 eV) is smaller than that of Bi (~1.25 eV) [41]. Meanwhile, the energy gap between the two $p_y$ Dirac bands induced by SOC is actually larger for Pb@H-Si(111) (0.65 eV) than that of Bi@H-Si(111) (0.5 eV), suggesting that Pb may be a better choice to achieve large-gap QSH states on the substrate. This is in sharp contrast with the Kane-Mele model in graphene, for which an energy gap is opened at Dirac point that is in proportion to the strength of SOC [19].

Besides Bi and Pb, we have also conducted calculations of other heavy elements adsorption on the Si surface, including Sb, Sn, Tl, In and Ga. It is found that Sb and Sn have a similar band structure with Bi and Sn, respectively (see Fig. S5 in Supplementary Information), but with a smaller energy gap resulting from their similar valence electron configurations but weaker SOC. Band structures of Tl, In and Ga@H-Si(111) are a bit different. As shown in Fig. 5, the Fermi level now sits further below the lower dispersive band that is mainly made of the heavy atom $p_y$ orbital. This is due to the one (two) less valence electron compared to the Pb (Bi) group, i.e., $[Xe].4f^{14}.5d^{10}.6s^2.6p^1$ for Tl. Clearly, one sees that from Ga to Tl, the SOC gap between the lower Dirac band and dispersive band increases dramatically, from around 0.1 eV (for Ga) to 0.5 eV for (Tl), confirming the dependence of energy gap ($\Delta E_2$) on SOC as demonstrated in Fig. 4.

In summary, we demonstrate the possibility of 'controlled' growth of large-gap topological quantum phases on conventional substrate surfaces such as the important Si surface by a unique approach of substrate orbital filtering process combined with a proper choice of SOC. Its underlying physical principles are general, applicable to deposition of different metal atoms on different substrates [11,27]. It opens up a new and exciting avenue for future design and fabrication of room temperature topological surface/interface states based on current available epitaxial growth and semiconductor technology, where the metal overlayer is atomically bonded but electronically isolated from the underneath semiconductor substrate [27].

**Methods**

Our electronic structure calculations based on density functional theory were performed by using a plane wave basis set [42] and the projector-augmented wave method [43], as implemented in the VASP code [44]. The exchange-correlation functional was treated with the generalized gradient approximation in Perdew-Burke-Ernzerhof format [45]. Calculations of $Z_2$ triviality were carried out by using the full-potential linearized augmented plane-wave method implemented in the WIEN2K package [46]. Details for models and computations are presented in Supplementary Information [28].

**Acknowledgements**

This research was supported by DOE (Grant No: DEFG02-04ER46148); Z. F. Wang and W. Ming additionally thank support from NSF-MRSEC (Grant No. DMR-1121252). We thank NERSC and the CHPC at University of Utah for providing the computing resources.

**Author contributions**

M. Z. carried out the theoretical calculations with the assistance of W. M. M., Z. L., Z. F. W. and Y. G. Y.; F. L. guided the overall project. M. Z. and F. L. wrote the manuscript.

**Additional information**

**Supplementary Information** accompanies this paper is available.

**Competing financial interests:** The authors declare no competing financial interests.

**Figure Legends**

Figure 1 A two-step approach to fabricate 2D TI by deposition of heavy metal atoms on a patterned H-Si(111) surface. (a) Schematic view. (b, c) The top and side view of the proposed structure, with the surface unit cell vector ($a_1$, $a_2$) indicated in (b) and the adsorption length $d$ in (c). (d) The first surface Brillouin zone.

Figure 2 Band structures of Bi and Pb@H-Si(111). (a-b) Without SOC. The Fermi level is set at zero. The green (yellow) shaded area represents the valence bands (conduction bands) of Si. Band compositions around Fermi level are also indicated. (c-d) Same as (a-b) with SOC.

Figure 3 Electronic structures of Bi@H-Si(111) and its edge state. (a) Comparison of band structures for Bi@H-Si(111) calculated by DFT (black lines) and Wannier function method (green circles). (b) The Dirac edge states within the SOC-induced band gap. Scale bar is indicated on the right. (c) The partial DOS projected onto $p_x$, $p_y$, and $p_z$ orbitals of Bi, and the total DOS of neighboring Si atoms. (d) Top: The charge density redistribution induced by metal atom surface adsorption for Bi@H-Si(111) (isovalue = 0.02 $e/\text{Å}^3$), illustrating saturation of Bi $p_z$ orbital. Bottom: Same as Top for the H-saturated freestanding planar Bi lattice.

Figure 4 Energy bands resulting from the four-band model [Eq. (1)] as a function of SOC strength ($\lambda$) scaled by $t$ ($t$ is the coupling strength between neighboring $p_x$ and $p_y$ orbitals). Fermi energy is set to zero. The SOC induced energy gaps ($\Delta E_1$, $\Delta E_2$ and $\Delta E_3$) are indicated. The global gap transition from $K$ point to $\Gamma$ point driven by SOC can be clearly seen.

Figure 5 Band structures of Ga, In, Tl@H-Si(111). (a-b) Band structures of Ga@H-Si(11) without and with SOC, respectively. The Fermi level is set to zero. Band compositions around Fermi level are indicated. (c-d) Same as (a-b) for In@H-Si(111). (e-f) Same as (a-b) for Tl@H-Si(111).

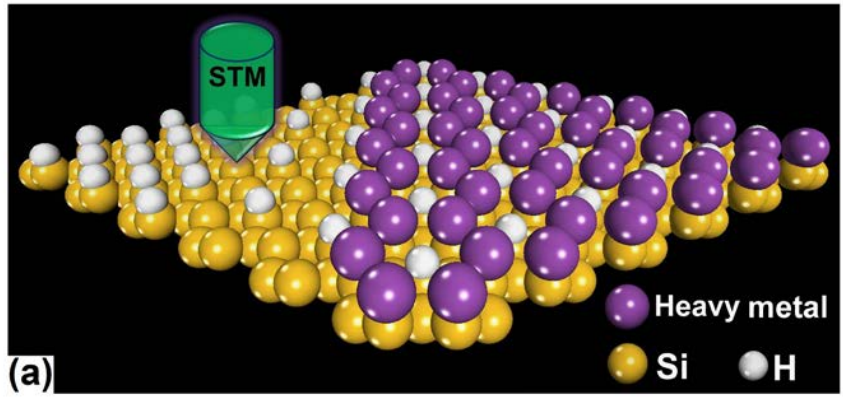

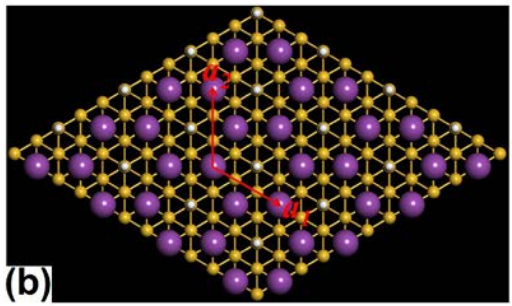
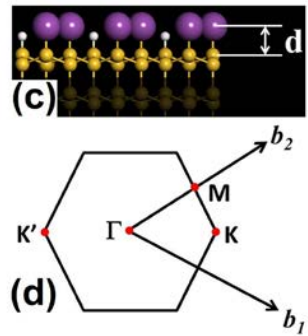

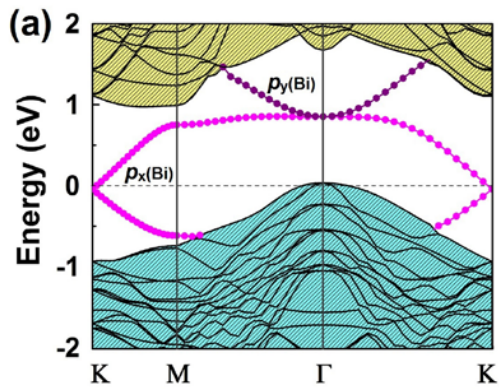
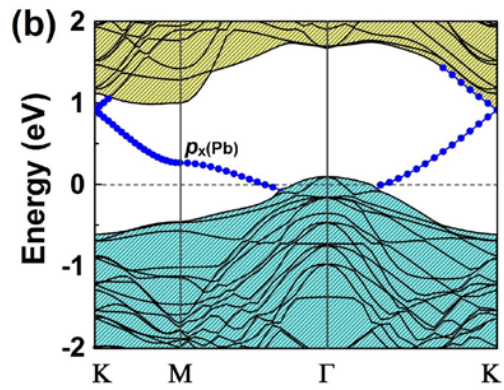

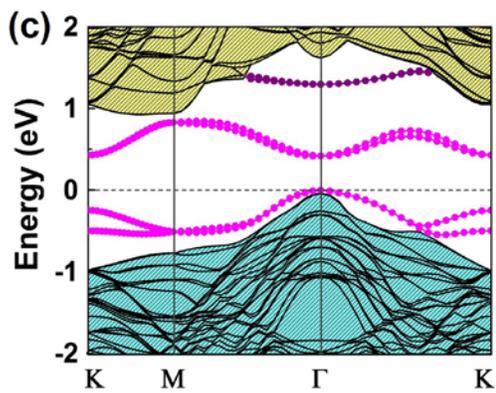
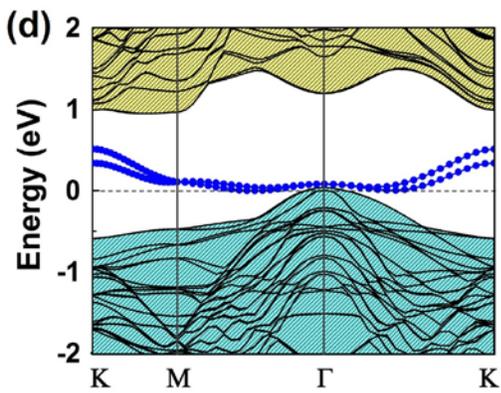

**Figs.1&2**

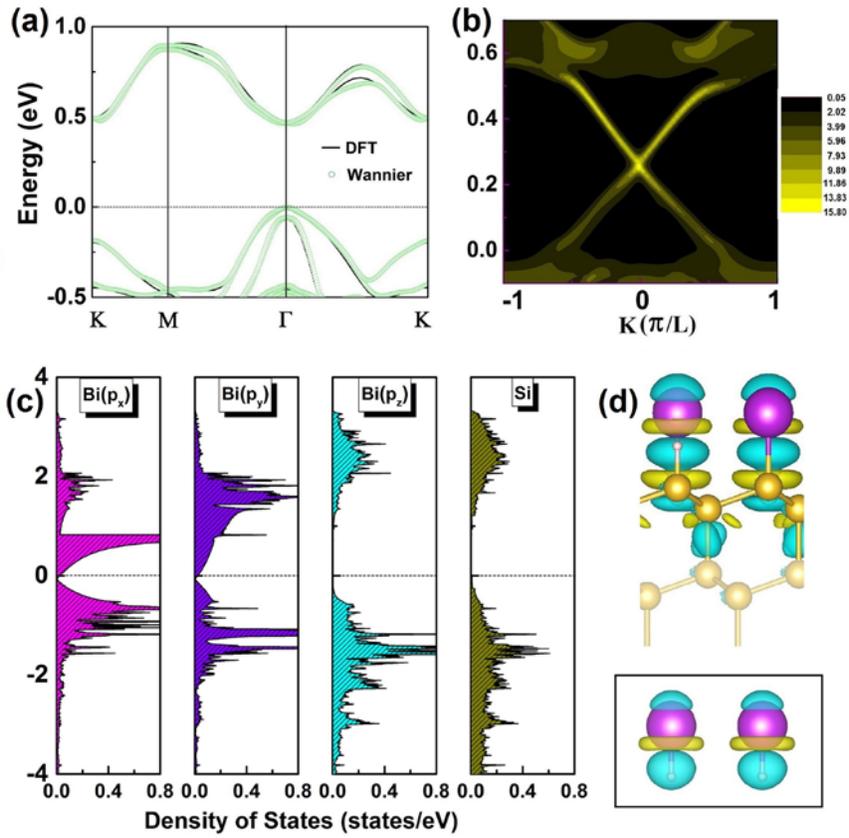

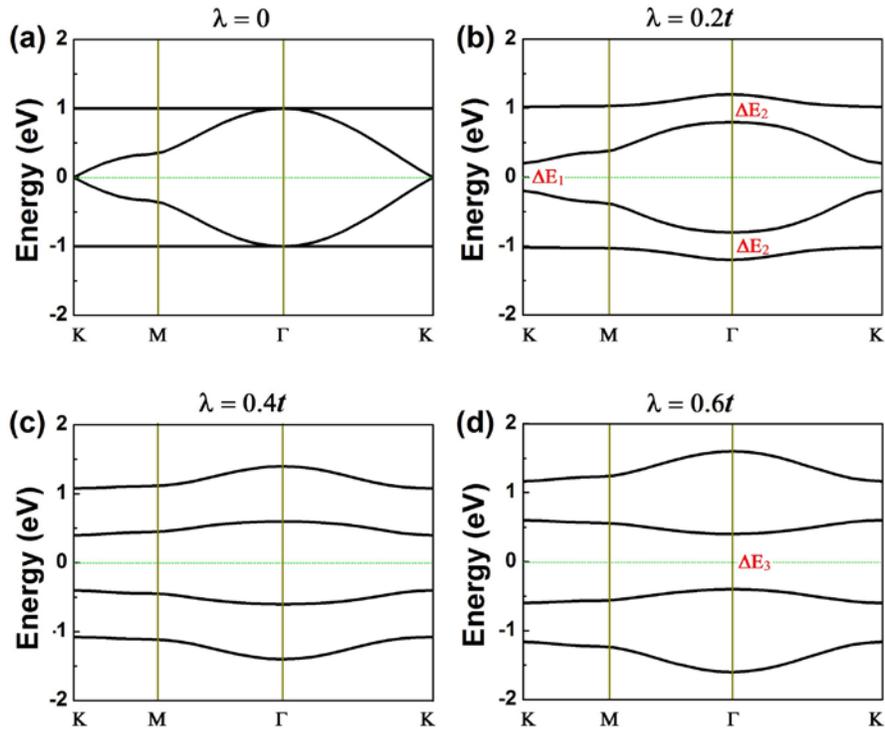

**Figs. 3&4**

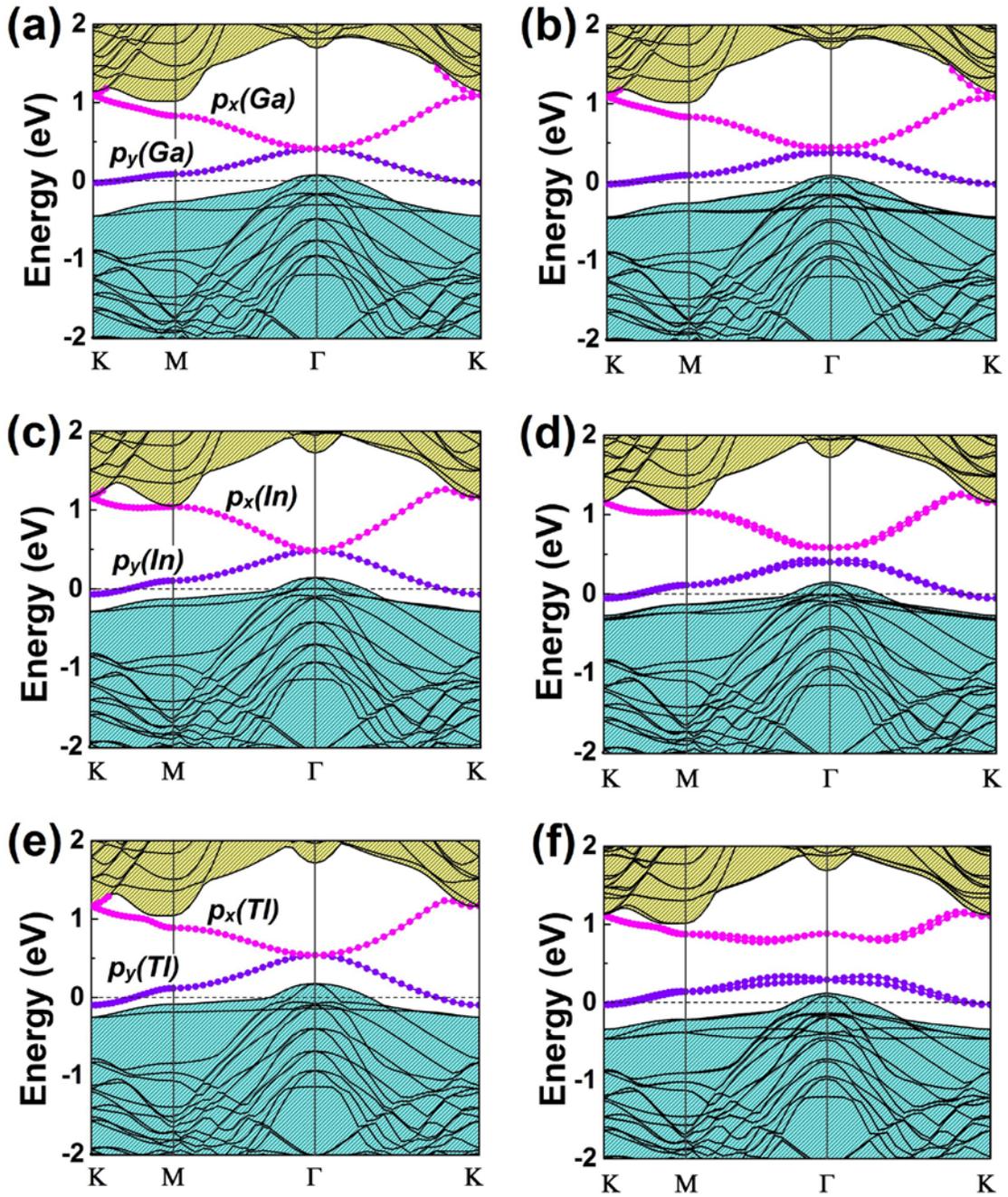

**Fig.5**

# -Supplementary Information-

# Formation of quantum spin Hall state on Si surface and energy gap scaling with strength of spin orbit coupling

*Miao Zhou[1], Wenmei Ming[1], Zheng Liu[1], Zhengfei Wang[1], Yugui Yao[2], and Feng Liu[1,3*]*

[1] Department of Materials Science and Engineering, University of Utah, UT 84112

[2] School of Physics, Beijing Institute of Technology, Beijing, China 100081

[3] Collaborative Innovation Center of Quantum Matter, Beijing, China 100084

*\*fliu@eng.utah.edu*

**CONTENTS**

## I. Computational Details

First-principles electronic structure calculations based on density functional theory (DFT) were carried out using the plane-wave-basis-set and the projector-augmented-wave method, as implemented in the VASP code. The energy cutoff was set to 500 eV. For the exchange and correlation functional, the generalized gradient approximation (GGA) in Perdew-Burke-Ernzerhof (PBE) format was used. Spin-orbit coupling (SOC) is included by a second variational procedure on a fully self-consistent basis.

H-Si(111) surfaces were modeled by using a slab geometry of ten atomic layers, with a vacuum region of 30 Å in the direction normal to the surface. Test calculations were performed by using larger thickness (twelve and sixteen layers) which gave similar results. In the pure H-Si(111) surface, both the top and bottom Si surfaces were terminated by H atoms in a monohydride form. For heavy metal atom (Bi, Pb, Sb, Sn, Ga, In and Tl) deposited H-Si(111), a $\sqrt{3}\times\sqrt{3}$ supercell was used with two of the three H atoms removed and re-adsorbed with heavy atoms into a hexagonal symmetry. During structural optimization, both the tenth layer of Si atoms and the H atoms saturating them were fixed and all other atoms were fully relaxed until the atomic forces were smaller than 0.01 eV/Å. A 15×15×1 $\Gamma$-centered $k$-point mesh was used to sample the Brillouin zone. Dipole corrections were also tested and found making little difference.

$Z_2$ invariant calculations were performed by using the program package WIEN2K. We employed the full-potential linearized augmented plane-wave method[2] within the GGA-PBE functional including SOC. A converged ground state was obtained using 5000 k-points in the first Brillouin zone and $K_{max} \times R_{MT} = 8.0$, where $K_{max}$ is the maximum size of the reciprocal lattice vectors and $R_{MT}$ denotes the muffin-tin radius. Wave functions and potentials inside the atomic sphere are expanded in spherical harmonics up to $l = 10$ and 4, respectively. For $Z_2$ calculation, we follow the method by Fukui *et al.*[2], to directly perform the lattice computation of the $Z_2$ invariants from first-principles, of which the detailed methodology is presented in the Refs. [21] and [36] of the paper. Here we present the calculated $Z_2$ number of Bi/Pb@H-Si(111), planar Bi hexagonal lattice, and Bi lattice with one side saturated by H, in Figs. S1-S4, respectively.

Furthermore, effective tight-binding (TB) Hamiltonians of the four-band model involving $p_x$ and $p_y$ orbitals in a hexagonal lattice with SOC are constructed (Eq. 1 in the paper) to qualitatively understand the SOC effect on the band gap opening mechanism. The results are shown in Fig. 4, from which we can see a gradual change of different gap sizes leading to a shift of global gap from $K$ to $\Gamma$ point, which correctly reproduces the DFT results of Pb@H-Si(111) and Bi@H-Si(111), as shown in Figs. 2(d) and (c) in the paper.

## II. $Z_2$ Invariant Calculation Results

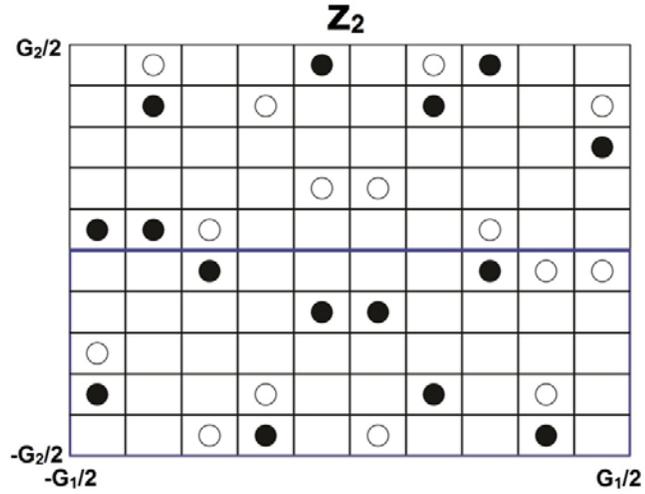

Fig. S1: The n-field configuration for Bi@H-Si(111). The calculated torus in Brillouin zone is spanned by $G_1$ and $G_2$ (Note that the two reciprocal lattice vectors form an angle of 120°). The solid dots and open circles denote n = 1 and −1, respectively, and the blank denotes 0. The $Z_2$ invariant is calculated by summing the n fields over half of the tori, which gives $Z_2 = 1$ indicating a topological insulator, in agreement with the edge state calculations [Fig. 3(b)].

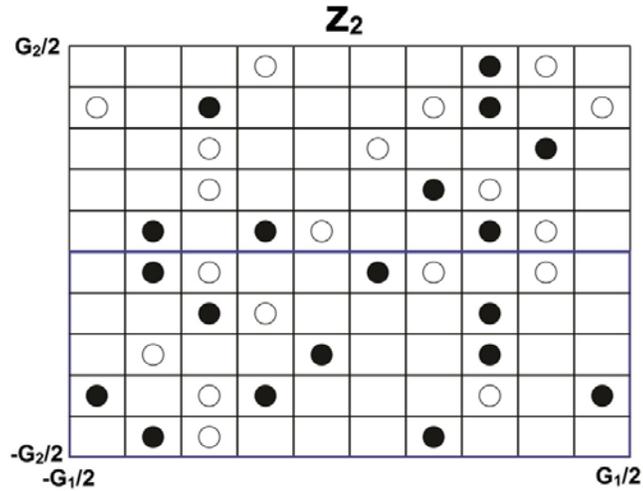

Fig. S2: Same as FIG. S1 for Pb@H-Si(111), by assuming a shift of Fermi level above the lower branch of Dirac band, as shown in Fig. 2 (e). $Z_2 = 1$, indicating a topological insulator.

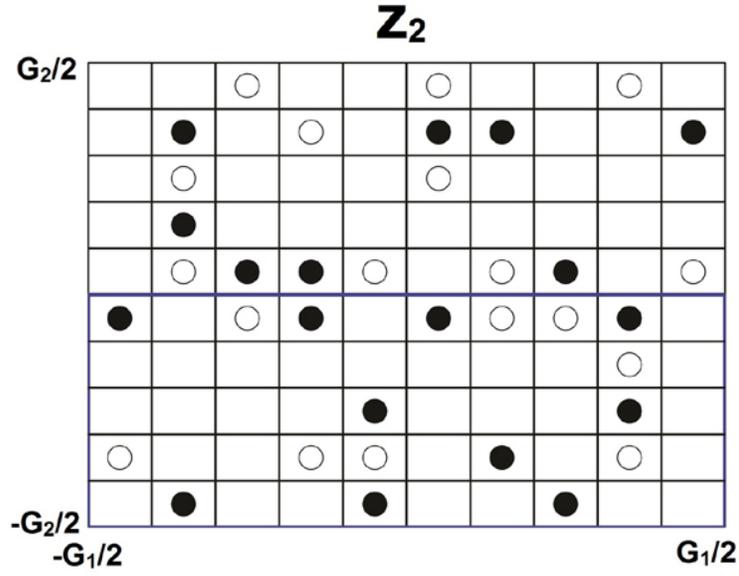

Fig. S3: Same as FIG. S1 for the planar hexagonal lattice of Bi. $Z_2 = 0$, indicating a trivial insulator.

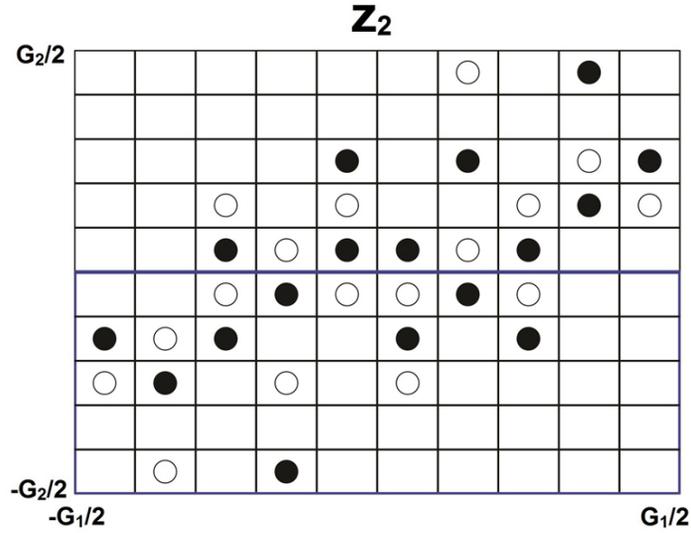

Fig. S4: Same as FIG. S1 for the planar Bi hexagonal lattice with one side saturated by H. $Z_2 = 1$, indicating a topological insulator

## III. Band structures of Sb and Sn@H-Si(111)

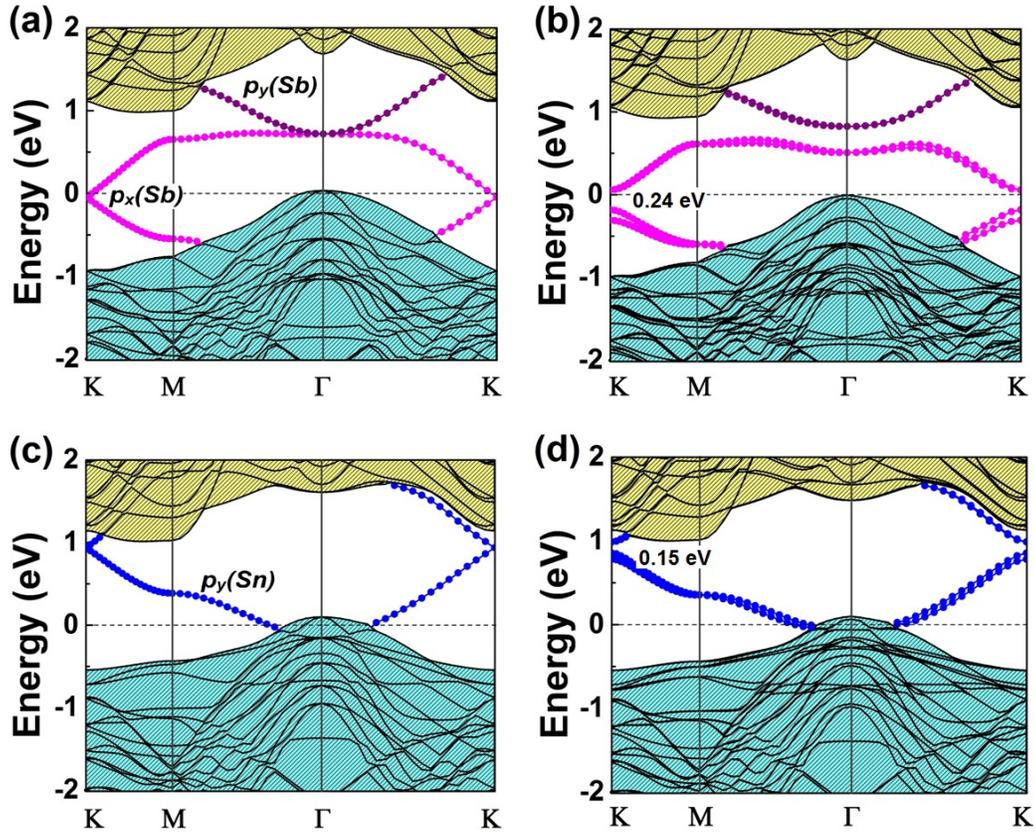

Fig. S5: (a-b) Band structures of Sb@H-Si(111) without and with SOC, respectively. The Fermi level is set to zero. (c-d) Same as (a-b) for Sn@H-Si(111). Band compositions around the Fermi level are indicated in (a) and (c), and the SOC induced energy gaps are indicated in (b) and (d).